\documentstyle[preprint,aps]{revtex}
\begin{document}
% \draft command makes pacs numbers print
\draft
\title{Shear Alignment and Instability\\ of Smectic Phases}
% repeat the author\address pair as needed
\author{Mark Goulian and Scott
T. Milner}
\address{Exxon Research and Engineering Company\\ Annandale,
New Jersey 08801}
\date{\today}
\maketitle
\begin{abstract}
We consider
the shear flow of well-aligned one-component smectic phases, such as
thermotropic smectics and lamellar diblock copolymers, below the critical
region. We show that, as a result of thermal fluctuations of the layers,
parallel ($c$) alignment is generically unstable and perpendicular ($a$)
alignment is stable against long-wavelength undulations. We also find,
surprisingly, that both $a$ and $c$ are stable for a narrow window of
values for the anisotropic viscosity.
\end{abstract}
% insert suggested PACSnumbers in braces on next line

% body of paper here
In the presence of simple shear flow, smectic phases
exhibit a surprising degree of complexity. As shear rate and temperature
are changed, a variety of transitions in orientation and morphology have
been observed. Although one might expect the liquid layers  to simply
slide over each other with their normals parallel to the shear gradient,
the $c$ orientation (Fig.\ \ref{orien}), they often orient with their
layer normals pointing in the vorticity (or neutral) direction, the $a$
orientation (Fig.\ \ref{orien}). This behavior is seen under some
conditions in both thermotropic smectics \cite{ssp,ssb,ish,par} and
lamellar phases of diblock copolymers \cite{ktb1,ktb2,wp1,wp2,kk}. For
thermotropics near the nematic-smectic transition, it has been shown that,
as a result of nematic fluctuations, $a$ alignment is favored over $c$
\cite{ssb}. In diblock copolymers, it has been shown that the wave vector
dependence of the quartic coupling in the Hamiltonian describing the
order-disorder transition favors $a$ \cite{gf}. In this paper we consider
the steady shear flow of generic one-component smectic phases at
temperatures well below the critical regime. We show that well-aligned
(i.e. defect free) systems favor the $a$ orientation: the $c$ orientation,
as well as orientations intermediate between $a$ and $c$, suffer an
instability from long-wavelength  undulations. We also find, surprisingly,
that within a small window of values for the anisotropic viscosity, both
$a$ and $c$ are stable.

It is clear that smectic phases will align so that the average flow
velocity has no component along the layer normals.  Otherwise, the layers
will be forced to deviate from their preferred spacing, which is
energetically costly. For perfectly flat layers, both $a$ and $c$
orientations permit steady shear flows with the layer displacement
unperturbed from its equilibrium value. However, thermal fluctuations of
the layers are convected differently in the two cases.  As we will see
below, convection leads to a greater suppression of thermal fluctuations
in the $c$ orientation than in the $a$ orientation. Hence, if we adopt a
naive picture in which the steady-state dynamics is determined by
maximizing layer fluctuations (free energy minimization), the $a$
orientation will be obtained in steady state. Since entropy maximization
arguments in non-equilibrium systems are often suspect, we also compute
the dynamic response function for small  perturbations away from an
aligned state and show that the $c$ orientation is indeed unstable towards
rotation to $a$.

Well below the ordering transition, where the amplitude of the order
parameter (concentration variations for diblocks or density variations for
thermotropics) is fixed, a well-aligned smectic is parametrized by a layer
displacement $u(r)$. We take the average layer normals parallel to $\hat
z$. The Hamiltonian for layer fluctuations is (in fourier space)\cite{dgp}
\begin{eqnarray} & &{\cal H}={1\over2}\int{\rm d}^3q\,u(-q)u(q)\omega(q)
\label{ham}\\ & &\omega(q)=Bq_z^2+Kq_\bot^4\nonumber \end{eqnarray} where
$B$ and $K$ are, respectively, the layer compression and bending moduli
and  $q_\bot$ denotes the component of the wave vector $\vec q$ in the
$(x,y)$-plane. The penetration length $\lambda\equiv\sqrt{K/B}$ is
typically of order the layer spacing. As a result, modes with $q_z=0$,
which correspond to layer undulations, are much softer than layer
compressions, which have $q_\bot=0$.

Consider now the effect of a steady shear with average flow velocity
parallel to  $\hat x$
 \begin{eqnarray}
 & &\vec v(\vec r\,)=(\vec r\cdot\hat n)\;\dot\gamma\,\hat x\label{vel}\\
 & &\hat n=\cos\theta\,\hat z+\sin\theta\,\hat y\nonumber.
 \end{eqnarray}
The velocity gradient direction is $\hat n$; in the $a$ orientation $\hat
n=\hat y$ and in the $c$ orientation $\hat n=\hat z$ (Fig.\ \ref{orien}).
A mode with wave vector $\vec q$ at time $t=0$ will be convected by the
flow Eq. (\ref{vel}) according to
 \begin{equation}
 \vec q(t)\equiv\vec q-\hat n\,\dot\gamma\,q_x\,t.
 \label{conv}
 \end{equation}

In the $a$ orientation, the $z$ component of $\vec q$ is unaffected by the
shear, $q_z(t)=q_z$; a mode that is a pure undulation ($q_z=0$) at $t=0$
remains a pure undulation at later times. In the $c$ orientation, however,
$q_z(t)=\dot\gamma\,q_x\,t$; undulations with $q_x\ne0$ at $t=0$ pick up a
compressional character ($q_z\ne0$) at later times. We will see below that
the lifetime for a fluctuation with wave vector $\vec q$ is given by
$1/(\omega(q)\beta(q))$ where $\beta(q)$ is a kinetic coefficient. As
noted above, layer compressions are much stiffer than undulations, i.e.
for fixed $|\vec q|$, $\omega(q)$ is larger for compression modes than for
undulations. Therefore, the lifetime of modes with $q_x\neq0$ is much
shorter in the $c$ orientation than in the $a$ orientation. (Here we
assume an isotropic kinetic coefficient for simplicity.) With a greater
suppression of modes in case $c$, we expect a corresponding increase in
the ``free energy'' and hence an instability towards rotation to $a$. (It
is easy to see that orientations intermediate between $a$ and $c$ suffer
from a similar suppression of modes, though to a lesser extent, and thus
the ``free energy'' should monotonically decrease in passing from $c$ to
$a$.)

In the previous discussion we have invoked free energy minimization in a
non-equilibrium system. One can often determine the steady-state behavior
in such cases by minimization of a quasi free energy. However, this
usually requires that the equations of motion  can be written in the
relaxational form $\dot\phi=\delta\Gamma/\delta\phi$, where $\phi$ is the
dynamical variable of interest and $\Gamma(\phi)$ is some functional. In
our case, one can consider the equation of motion for the angle $\theta$
parametrizing orientations between $a$ and $c$ (Fig.\ \ref{orien}) after
averaging over layer fluctuations. Unfortunately, we have not been able to
express the equations of motion in a form that would justify a
minimization principle. However, the above picture suggests an alternative
course. A long-wavelength undulation in the $\hat y$ direction locally
looks like a change in $\theta$, i.e. a tilt of the layers in the neutral
direction. Therefore, a local driving force for rotating from $c$ to $a$
should appear in the $c$ orientation as an instability  towards
long-wavelength undulations with $\vec q\,||\,\hat y$.

Below, we first compute a quasi free energy to demonstrate the mode
suppression described above. We then compute the dynamic response function
(to one loop in perturbation theory) and show that indeed the $c$
orientation, as well as orientations intermediate between $a$ and $c$, are
unstable and the $a$ orientation is stable.

To describe the hydrodynamics of smectics, we follow the treatment of
\cite{mpp,dgp}. We neglect inertial terms and assume incompressibility. In
this limit, the dynamics of $u$ reduces to the relaxational form:
\begin{equation} {\partial u(q)\over\partial t}+\beta(q)\,{\delta {\cal
H}\over\delta u}=0,\label{hydro} \end{equation}
 \begin{equation}
 \beta(q)=b_p+{q_\bot^2\over\eta q^4+\eta' q_z^2q_\bot^2}.
 \end{equation}
$\eta$ and $\eta'$ are viscosities \cite{visc}, and $b_p$ is the
permeation constant. Although our results are independent of $b_p,$ we
include permeation in order to ensure that subsequent expressions converge
for large $q_z$.

In the presence of steady shear,  the time derivative in Eq.\
(\ref{hydro}) is replaced by the convective derivative $\partial/\partial
t-\dot\gamma q_x\partial/ \partial q_n$, $q_n\equiv\vec q\cdot\hat n$. In
order to take into account thermal fluctuations, we add to the right hand
side of Eq.\ (\ref{hydro}) a random noise, $\zeta(q,t)$, with correlations
that insure for zero shear rate the system relaxes to equilibrium
\cite{tas}. We thus have
 \begin{equation}
 ({\partial\over\partial t}-\dot\gamma q_x{\partial\over\partial
 q_n})u(q)+\beta(q)\omega(q) u(q)=\zeta(q,t)\label{eqm}
 \end{equation}
 \begin{equation}
 \langle\zeta(q,t)\zeta(-q,0)\rangle=2k_BT\,\beta(q)\delta(t),\nonumber
 \end{equation}
where $k_B$ is Boltzmann's constant and $T$ is temperature.
{}From Eq.\ (\ref{eqm}), we find for the equal-time correlation function
of $u$:
 \begin{eqnarray}
 \chi(q)\equiv\langle u(q,0)&&u(-q,0)\rangle
 =k_BT\int_{-\infty}^0{\rm d}t\,2\beta(q(t))\nonumber\\ &
 &\times\exp\left(-2\int_{t}^0{\rm d}t' \beta(q(t'))\omega(q(t'))\right).
 \end{eqnarray}
Similar expressions are found in \cite{ok,fl,ram,br}.

Since the equation of motion Eq.\ (\ref{eqm}) is linear, $\chi$ determines
the probability distribution of $u$. We thus define a ``free energy''
density by
 \begin{equation}
 {\cal F}\equiv-{k_BT\over2}\int{\rm d}^3q\;\log\,\chi(q).
 \label{fren}
 \end{equation}
 Eq.\ (\ref{fren}) is a complicated function of $\dot\gamma$ and $\theta$,
however we can extract
the leading asymptotic behavior as $\dot{\bar\gamma}\to0,$ where
$\dot{\bar\gamma}\equiv\dot\gamma\eta/B$. For thermotropics, with typical
values of $B=10^8\,{\rm erg/cm^3}$ and $\eta=1\,{\rm poise}$ \cite{dgp},
small $\dot{\bar\gamma}$ implies shear rates $\dot\gamma<10^{8}\,{\rm
s}^{-1}$, which easily encompass the range studied experimentally. For
block copolymers, however, $B$ can be much smaller and $\eta$ much larger
by many orders of magnitude. For example, for $B=10^6\,{\rm erg/cm^3}$
\cite{ah} and $\eta=10^6\,{\rm poise}$ \cite{ktb1} we must take
$\dot\gamma<1\,{\rm s}^{-1}$.  We find for small $\dot{\bar\gamma}$
 \begin{equation}
 {\cal F}={\cal F}|_{\dot{\bar\gamma}=0}+c_1{k_BT\over\lambda^3}\,({\dot
 {\bar\gamma}}\cos\theta)^{4/3}+\ldots,
\end{equation}
where $c_1\approx 0.25$ is a constant determined numerically. ${\cal F}$
will therefore be minimized for $\theta=\pi/2$, which is the $a$
orientation.

While the above analysis is intuitively appealing in its treatment of
thermal fluctuations, as we have discussed in the introduction, it is not
well justified.  We therefore turn to a computation of the dynamic
response function.

We can model a time dependent disturbance to the system by shifting the
noise in Eq.\ (\ref{eqm}) $\zeta(q,t)\to\zeta(q,t)+f(q,t).$  The response
of the system to $f(q,t)$ is given by
 \begin{eqnarray}
 & &\langle u(q,t)\rangle=\int_{-\infty}^{t}{\rm d} t'\,C(q,t,t')
 f(q(t'-t),t')\label{lnrsp}\\ & &C(q,t,t')=\exp\left(-\int_{t'-t}^0{\rm
 d}t''\, \beta(q(t''))\omega(q(t''))\right).\label{rspfn}
 \end{eqnarray}
For all orientations $\theta$, disturbances created by $f$ relax
exponentially with a decay rate that depends on the convected value of
$\vec q$. In order to search for an instability, we must go beyond the
linear Eq.\ (\ref{eqm}). The smectic Hamiltonian with the leading
anharmonic corrections is given by \cite{gp}
 \begin{eqnarray}
 {\cal H}=\int{\rm d}^3r\,\big[{B\over2}(\partial_z
 u&+&{1\over2}(\nabla_\bot u)^2)^2 + {K\over2}(\nabla_\bot^2u)^2
 \nonumber\\ &+&C(\partial_zu+{1\over2}(\nabla_\bot
 u)^2)\big].\label{anhar}
 \end{eqnarray}
The last term in Eq.\
(\ref{anhar}) is a counterterm that is used to enforce  the condition
\cite{gp}
 \begin{equation}
 \langle{\partial u\over\partial z}\rangle|_{f=0}=0,\label{dudz}
 \end{equation}
which ensures that $u$ describes deviations from the average layer spacing.

Although we have had to include nonlinear corrections in ${\cal H}$, this
does not mean that the physics underlying the response function  differs
from the physics presented in the introduction and contained in ${\cal F}$,
which was only computed for the quadratic Hamiltonian Eq.\ (\ref{ham}).
The non-linear terms in Eq.\ (\ref{anhar}) are not arbitrary: they are
determined by requiring that  rotating the layers (in the absence of
shear) does not cost any energy \cite{gp}. Since ${\cal F}$ has been
calculated for arbitrary tilt $\theta$ within the harmonic theory, it in
principle contains the same information as the anharmonic terms to one
loop.

The equations of motion are now nonlinear
 \begin{equation}
 ({\partial\over\partial t}-\dot\gamma q_x{\partial\over\partial
 q_n})u(q)+\beta(q)\,{\delta {\cal H}\over\delta u}=\zeta(q,t)+f(q,t),
 \label{aneqm}
 \end{equation}
but we are free to choose $f(q,t)$
arbitrarily small so that  linear response is still correct. For zero
shear $\dot{\bar \gamma}=0$, the response  function, to one loop order,
takes the form
 \begin{eqnarray}
 &&C(q,t,t')=\nonumber\\
 &&\exp\left(-\int_{t'-t}^0{\rm d}t''\,\beta(q(t''))
 (\omega(q(t''))+\delta\omega(q(t'')))\right).\label{rspfn2}
 \end{eqnarray}
The shift $\delta\omega(q)$ corresponds to a
renormalization of $B$ and $K$; rotation invariance forbids a term
proportional to $q_{\bot}^2$ from appearing \cite{gp}. In the presence of
shear, however, we can no longer invoke rotation invariance and we expect
terms quadratic in $q_x$ and $q_y$ to appear. For sufficiently long times,
the $q_x$ dependent terms will be small compared with the $Bq_z^2$ and
$Kq_\bot^4$ terms in $\omega(q(t))$. ($q_z$ and $q_y$ pick up $q_x$
dependence  through convection, Eq.\ (\ref{conv})). A contribution to
$\delta\omega$ that is proportional to $q_y^2$ and negative, on the other
hand, will be destabilizing. As discussed in the introduction, this is in
accord with the intuition that a local tilt of the smectic layers in the
neutral direction, i.e. a local change in $\theta$, corresponds to a mode
with $\vec q\,||\,\hat y$.

We have computed the response function $C(q,t,t')$ for Eq.\ (\ref{aneqm})
perturbatively to one-loop order. If we assume $f(q,t)$ is nonzero only
for $\vec q=q_y\hat y$, then, in the limit of long times (low
frequencies), linear response still takes the form of Eq.\ (\ref{lnrsp})
with the response function as in Eq.\ (\ref{rspfn2}). After taking care to
maintain Eq.\ (\ref{dudz}), we find, again in the limit of small
$\dot{\bar\gamma}$,
 \begin{equation}
 \delta\omega(q)\approx-c_2\,q_y^2\,(2+{\eta'\over\eta})
 {k_BT\over\lambda^3}
 (\dot{\bar\gamma}\cos\theta)^{4/3}+\ldots,\label{domga}
 \end{equation}
where $c_2\approx1.4\times10^{-3}$ is determined numerically. For $\vec q$
along $\hat y$, $\omega(q)$ goes as $q_y^4$ and so for sufficiently small
$q_y$, $\delta\omega>>\omega$. Hence, for orientations with $\theta$ not
too close to $\pi/2$ (i.e. away from the $a$ orientation), disturbances
with  $\vec q\,||\,\hat y$ will grow and the smectic will be unstable.
Here we have implicitly assumed $2+\eta'/\eta>0$; we will return to this
point below.

When $\theta$ is sufficiently close to $\pi/2$ such that
$\cos\theta\sim\dot{\bar\gamma}\sin\theta$, subleading terms in Eq.\
(\ref{domga}) will be important and can change the sign of $\delta\omega$.
To check the stability of the smectic in this limit, we have computed the
response function for  $\theta=\pi/2$. In this case we find
 \begin{equation}
 \delta\omega(q)\approx c_3\,q_y^2\,{k_BT\over\lambda^3}
 \dot{\bar\gamma}+\ldots,
 \end{equation}
where $c_3\approx4.6\times10^{-3}$. Thus, for the $a$ orientation, the
correction $\delta\omega$ is positive and the smectic is stable.

As was mentioned previously, in concluding that the $c$ orientation is
unstable we have assumed $2+\eta'/\eta>0$. In fact, there is a small
window in which the sign may be reversed. The five viscosities that enter
the constitutive relation between stress and strain rate in a uniaxial
fluid are constrained by the requirement of non-negative energy
dissipation \cite{mpp}. In our case \cite{visc}, this requires $\eta\geq0$
and $\eta'\geq-4\eta$. Thus, as $\eta'$ is lowered into the range
$-4\eta\leq\eta'\leq-2\eta$, we find a transition to a regime in which
the  smectic is stable for all orientations $\theta$. This range of
viscosities corresponds to small dissipation for uniaxial extensional flow
along $\hat z$ compared with the dissipation for shear flow.
Unfortunately, we have no simple physical argument for this result.

We have considered the shear flow of well-aligned one-component smectic
phases outside of the critical regime. Very close to the nematic-smectic
transition, where nematic fluctuations are large, our analysis is
inappropriate. However, the appearance of the $a$ orientation in this
regime has been accounted for in \cite{ssb}. Our work is similarly
complementary to that in \cite{gf}, where the role of amplitude
fluctuations is considered. Also note that our analysis requires
modification for two-component systems  (i.e. lyotropic smectics) where
there is an additional hydrodynamic variable.

We have argued that, as a result of convection and the higher energetic
cost of compressing smectic layers compared with bending, there is a
greater suppression of fluctuations in the $c$ orientation. With the naive
view that the steady-state behavior is determined by minimization of a
``non-equilibrium free energy'', the $a$ orientation will be favored over
$c$. To demonstrate the scenario suggested by these arguments, we have
computed the dynamic response of the  system to a long-wavelength
perturbation corresponding to a local tilt of the layers. We find that
over most of the range of allowed values for the anisotropic viscosity,
the $c$ orientation and orientations intermediate between $c$ and $a$ are
indeed unstable and the $a$ orientation is stable. Surprisingly, we have
also found that there is a window of values for the viscosity in which all
orientations are stable. Our treatment completely neglects the role of
defects as well as the possibility of a nonlinear  relation between stress
and strain rate (non-Newtonian behavior), both of which are likely to play
an important role in some ranges of temperature and shear.  While our
analysis by no means accounts for the entire phase behavior of smectics
under shear, we suggest that the mechanism described in this paper may
account for the prevalence of the $a$ orientation observed in the shear
flow of one-component smectics.

\acknowledgments

We would like to thank Glenn Fredrickson for helpful discussions.

%now thereferences..

% figures follow here
%
\begin{figure}
\caption{Schematic of the $c$, $a$,
and intermediate orientations. We always take the flow velocity to be in
the $\hat x$ direction and the layer normals to point in the $\hat z$
direction. The shear plane is in the plane of the page.}
\label{orien}
\end{figure}

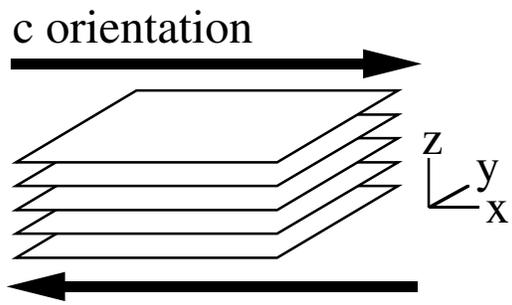
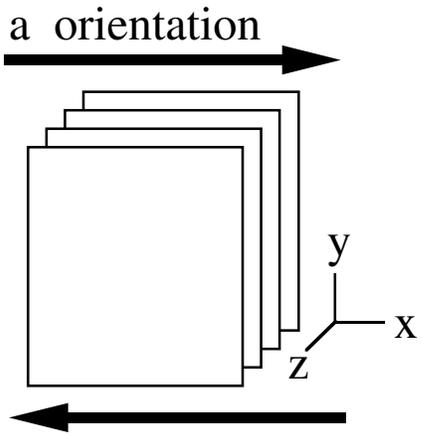
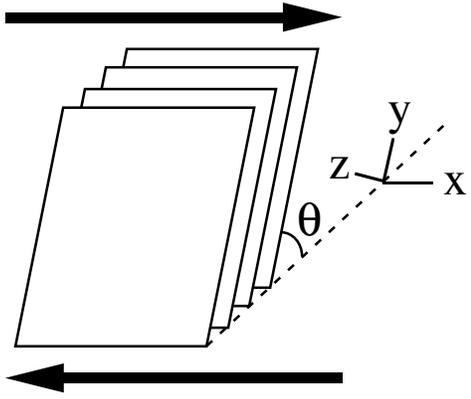
\end{document}